\PassOptionsToClass{twocolumn}{revtex4-1}

\documentclass[%
  aps,
  10pt,
  floatfix,%
  prl,%
  superscriptaddress,%
  a4paper,%
  showpacs,%
]{revtex4-1}

\usepackage[T1]{fontenc}
\usepackage[latin1]{inputenc}
\usepackage[english]{babel}

\usepackage{textcomp}
\usepackage{amsmath}
\usepackage{amsfonts}
\usepackage{amssymb}

\usepackage{units}
\usepackage{braket}

\usepackage{xcolor}
\usepackage{graphicx}
\usepackage{subfigure}
\usepackage[textfontcmd=small,vecinclude=overwrite]{combinedgraphics}

 \usepackage{ifpdf}
 \ifpdf
   \newcommand*{\hyperrefcolor}{violet}
   \usepackage[%
     pdfsubject={Preprint},%
     pdftitle={Single ions trapped in a one-dimensional optical lattice},%
     pdfauthor={M. Enderlein, T. Huber, Ch. Schneider, and T. Schaetz},%
     bookmarksopen=true,   %% open bookmark tree
     plainpages=false,     %% simple, arabic page number anchors
     colorlinks,           %% colored links instead of frames
     linkcolor=\hyperrefcolor,%
     anchorcolor=\hyperrefcolor,%
     citecolor=\hyperrefcolor,%
     urlcolor=\hyperrefcolor,%
     menucolor=\hyperrefcolor,%
     filecolor=\hyperrefcolor%
   ]{hyperref}
 \fi

\usepackage{userdef_macros}

\def\mybibstyle{apsrev4-1}
\def\mypicturescale{1}
\def\mypicturescalepre{0.5}
\def\myfigurescale{0.677}  %% 5 inches to 86 mm
\def\myfigurescalepre{0.775}

\newcommand{\theintroductorynote}{%
  \pagenumbering{roman}
  \thispagestyle{empty}
  \section*{Introductory Note}
  This version of the manuscript contains the differences between the
  previous revision and the current one. Any change at each page is preceeded
  by a unique mark, \eg \changecntbox{1}. Changes between the revisions
  are emphasized like this: {\removedcolor\removedfont{text removed
  from the previous revision}} and {\addedcolor\addedfont{newly added text
  in the current revision}}. A list with comments explaining each change
  is appended at the end of the manuscript (see section \revisionchangesname).
  \clearpage~
  \thispagestyle{empty}
  \clearpage
}

\newcommand{\myfigureschematic}{
  \begin{figure}[!tb]
    \centering
    \includegraphics[width=\mypicturescale\hsize,angle=0,keepaspectratio]%
      {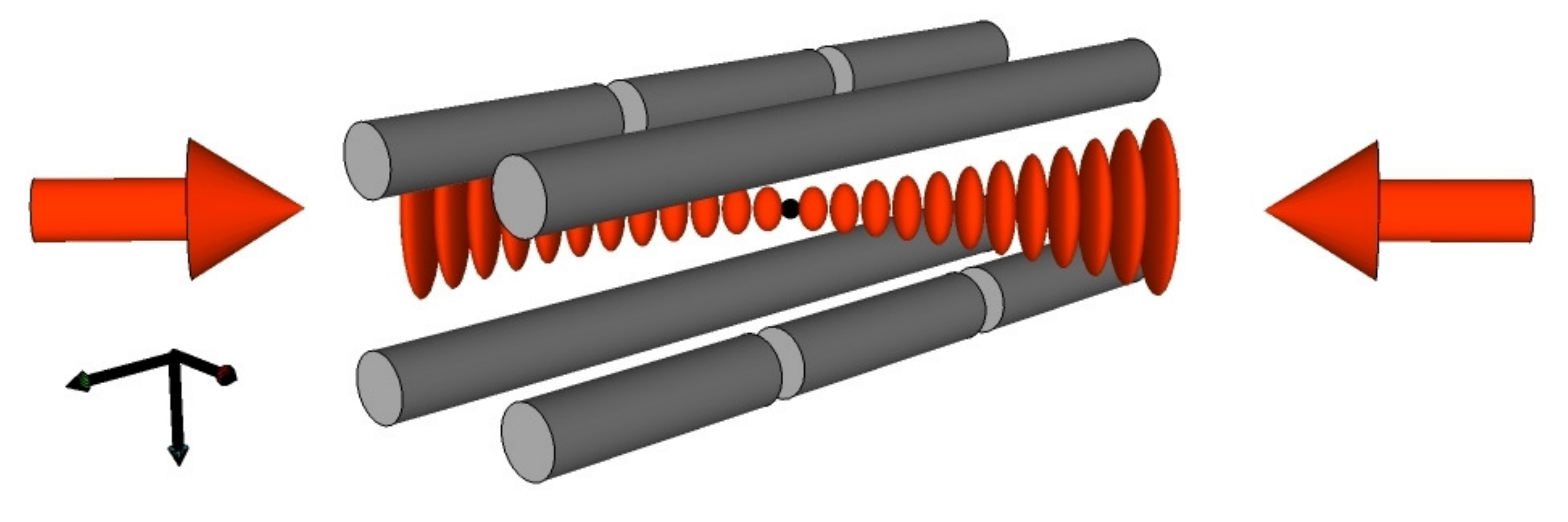}%
    \unitlength=0.01\hsize%
    \unitlength=\mypicturescale\unitlength%
    \begin{picture}(0,0)(100,0)
      \put(4,23.5){$\sigma_1^+$}
      \put(92,23.5){$\sigma_2^+$}
      \put(23,2){rf}
      \put(32,-1){dc}
      \put(8,3.5){\makebox(0,0)[lb]{x}}
      \put(17,6.5){\makebox(0,0)[rc]{y}}
      \put(5,11.5){\makebox(0,0)[tc]{z}}
    \end{picture}
    \caption{Schematic of the setup.
      The linear rf trap consists of four electrode rods. The rf voltage is
      applied to two rods while the others (segmented) remain at rf
      ground. This generates a two-dimensional confinement in the radial
      ($x$/$y$) directions. Additional dc voltages applied to the outer
      electrode segments add the axial confinement ($z$ direction).
      The counter-propagating dipole trap beams (arrows) propagate in the
      $y$-$z$ plane, crossing the $z$ axis at an angle
      of $\unit[45]{\degree}$. They are focussed on the ion at the center of
      the rf and dc potentials (black dot) and have waist radii of
      $w_0\approx\unit[5]{\micro m}$, and equal intensities. 
      The non-interfering configuration $\{\sigma_{1}^+\text{,}\sigma_{2}^-\}$
      leads to a Gaussian-shaped dipole potential (not shown) with twice the
      single-beam intensity. 
      For identical polarizations, $\{\sigma_{1}^+\text{,}\sigma_{2}^+\}$, the
      two beams interfere and form an additional standing-wave pattern in the
      direction of beam propagation, as shown.
      At the positions of constructive interference, the maximal intensity
      ideally is four times that of the single beam. 
      Not shown are additional Doppler cooling beams (propagating in the
      $x$-$z$ plane at an angle of $\unit[22.5]{\degree}$ to the $z$ axis).
      Doppler cooling fluorescence light from the ion is detected with a CCD
      camera above the trap.
    }
    \figlabel{setup}
  \end{figure}
}

\newcommand{\myfiguress}{
  \begin{figure}[!tb]
    \centering
    \includegraphics[width=\mypicturescale\hsize,keepaspectratio]%
      {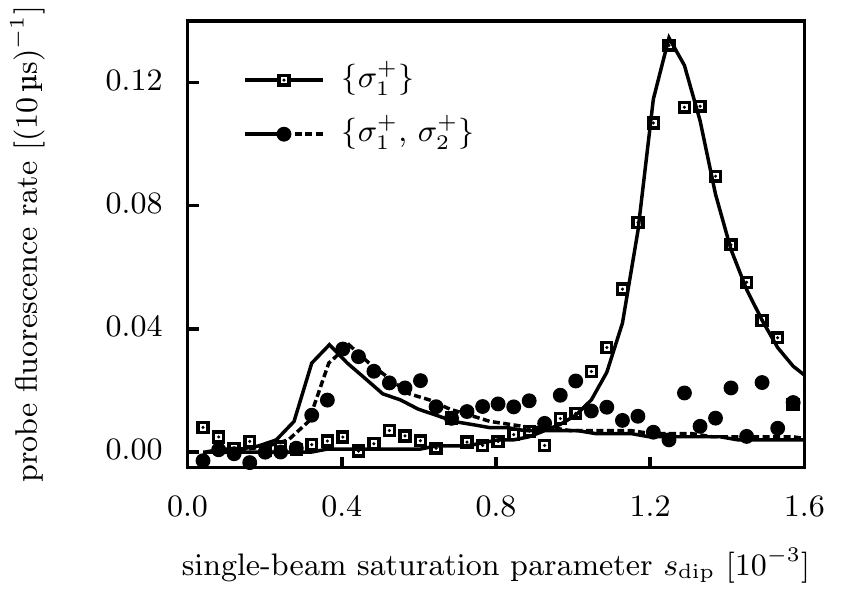}%
    \unitlength=0.01\hsize%
    \unitlength=\mypicturescale\unitlength%
    \caption{Measurement of the light shift induced by the dipole trap
      beam(s) on an ion stored in the rf trap. The fluorescence of an ion due to
      a weak, blue-detuned probe laser is measured as a function of the
      single-beam saturation parameter $s_{\text{dip}}$ of the dipole
      trap beam(s). The two configurations of the dipole trap beams are either
      a single Gaussian beam, $\{\sigma_{1}^+\}$, or a combination of two
      counter-propagating beams: $\{\sigma_{1}^+\text{,}\sigma_{2}^+\}$.
      The detection time is $\unit[10]{\micro s}$. 
      Each data point is the average of $4000$ measurements.
      Statistical errors of the fluorescence measurements are
      small compared to the size of the symbols.
      Statistical errors due to the subtraction of stray light, measured
      without ion, lead to an increasing variance for increasing
      $s_{\text{dip}}$ and are not considered in the error budget.
      The systematic error of $s_{\text{dip}}$ corresponds to
      an uncertainty of the absolute values of the $s_{\text{dip}}$ scale, which
      is not relevant here.
      Solid lines show the results of MCSs with $T_0=\unit[4]{mK}$. 
      The dashed line is a fit of the MCS to the measured resonance in the
      counter-propagating configuration. From this we derive an
      estimate for experimental imperfections requiring the higher
      $s_{\text{dip}}$ to shift the transition into resonance with the
      probe laser.
    }
    \figlabel{stark}
  \end{figure}
}

\newcommand{\myfiguretrapping}{
  \begin{figure}[!tb]
    \centering
    \includegraphics[width=\mypicturescale\hsize,keepaspectratio]%
      {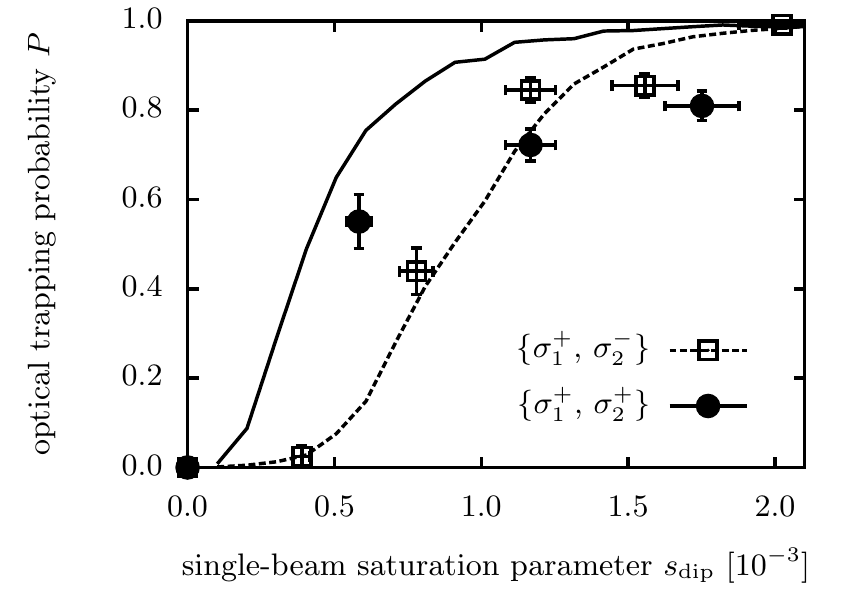}%
    \unitlength=0.01\hsize%
    \unitlength=\mypicturescale\unitlength%
    \caption{Optical trapping probability as a function of the single-beam
      saturation parameter for the non-interfering,
      $\{\sigma_{1}^+\text{,}\sigma_{2}^-\}$, and the interfering,
      $\{\sigma_{1}^+\text{, }\sigma_{2}^+\}$, configuration of the two
      counter-propagating dipole trap beams. The optical trapping time is  
      $T_{\text{opt}}=\unit[25]{\micro s}$.
      Data points represent the mean number of successful trapping attempts
      for typically $30$ ions with $1\sigma$ statistical errors for the
      trapping probability and systematic errors for the saturation intensity
      ($\delta P_{\text{dip}}/P_{\text{dip}}=\pm 0.03$, 
      $\sigma_{w_{0}}\approx\unit[0.27]{\micro m}$).
      Lines represent MCS results based on the ponderomotive approximation to
      the rf potential. Input parameters are $T_0=\unit[4]{mK}$, a power
      scaling factor of 0.88 and an offset force 
      $F=\unit[0.5\times 10^{-20}]{N}$ (see also footnote [31]).
    }
    \figlabel{trapping1}
  \end{figure}
}

\newcommand{\myfiguretrappingrf}{
  \begin{figure}[!tb]
    \centering
    \includegraphics[width=\mypicturescale\hsize,keepaspectratio]%
      {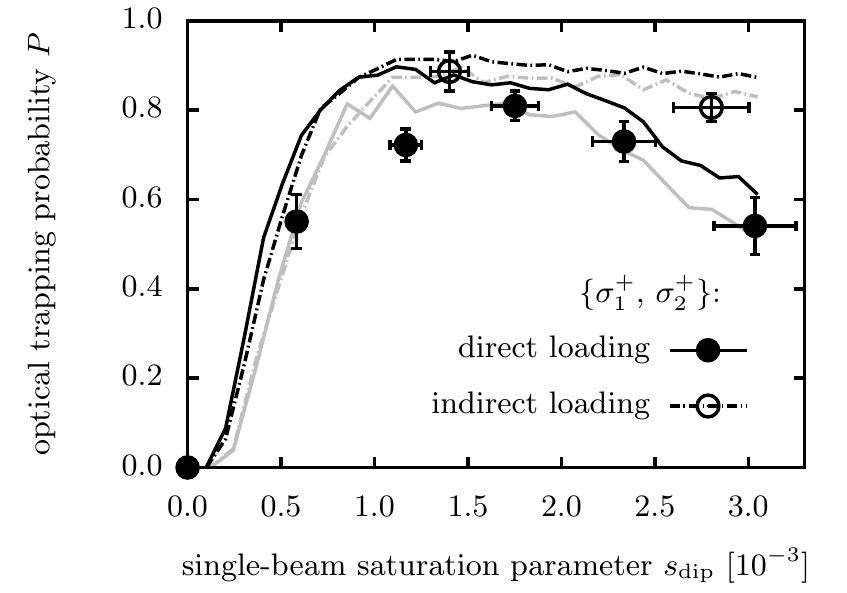}%
    \unitlength=0.01\hsize%
    \unitlength=\mypicturescale\unitlength%
    \caption{Optical trapping probability as a function of the single-beam
      saturation parameter for the interfering configuration,
      $\{\sigma_{1}^+\text{, }\sigma_{2}^+\}$. Comparison of two different
      protocols for loading of the optical lattice: Direct transfer from the rf
      trap to the optical lattice (as in \figref{trapping1},
      $T_{\text{opt}}=\unit[25]{\micro s}$) and indirect loading via an
      intermediate phase of trapping in a single-beam trap
      ($T_{\text{opt}}=\unit[100]{\micro s}$). Data points represent the mean
      number of successful trapping attempts for typically $30$ ions with
      $1\sigma$ statistical errors for the
      trapping probability and systematic errors for the saturation intensity
      (see \figref{trapping1}).
      Lines show MCS results for both loading sequences, taking into account the
      time-dependent rf potential (input parameters as in \figref{trapping1}).
      Gray lines represent MCS results that consider additional experimental
      imperfections.
    }
    \figlabel{trapping2}
  \end{figure}
}

\begin{document}

\normalsize 

%% smaller pictures in two-column layout
\iftwocolumn{}{\gdef\myfigurescale{\myfigurescalepre}}
\iftwocolumn{}{\gdef\mypicturescale{\mypicturescalepre}}

%% resets page numbers (useful if introductory note is typeset)
\pagenumbering{arabic}

\title{Single ions trapped in a one-dimensional optical lattice}

\def\myaffilb{%
\affiliation{%
  Max-Planck-Institut f\"{u}r Quantenoptik,
  Hans-Kopfermann-Str.~1,
  85748 Garching,
  Germany%
}}
\def\myaffila{%
\affiliation{%
  Albert-Ludwigs-Universit\"{a}t Freiburg,
  Physikalisches Institut,
  Hermann-Herder-Str.~3,
  79104 Freiburg,
  Germany%
}}

\author{Martin~Enderlein}
\author{Thomas~Huber}
\author{Christian~Schneider}
\author{Tobias~Schaetz}
\email{tobias.schaetz@physik.uni-freiburg.de}
\myaffila
\myaffilb


\begin{abstract}
  We report on three-dimensional optical trapping of single ions in an optical
  lattice formed by two counter-propagating laser beams. We characterize
  the trapping parameters of the standing wave using the ion as a
  sensor stored in a hybrid trap consisting of a radio-frequency
  (rf), a dc, and the optical potential.
  When loading ions directly from the rf into the standing-wave
  trap, we observe a dominant heating rate. Monte Carlo simulations confirm
  rf-induced parametric excitations within the deep optical lattice as the main
  source. We demonstrate a way around this effect by an alternative transfer
  protocol which involves an intermediate step of optical confinement in a
  single-beam trap avoiding the temporal overlap of the standing wave and the
  rf field.
  Implications arise for hybrid (rf/optical) and pure optical traps as
  platforms for ultra-cold chemistry experiments exploring
  atom--ion collisions or  quantum simulation experiments with ions, or
  combinations of ions and atoms.
\end{abstract}

\maketitle

%%%%%%%%%%%%%%%%%%%%%%%%%%%%%%%%%%%%%%%%%%%%%%%%%%%%%%%%%%%%%%%%%%%%%%%
%% document
%%%%%%%%%%%%%%%%%%%%%%%%%%%%%%%%%%%%%%%%%%%%%%%%%%%%%%%%%%%%%%%%%%%%%%%

%\section*{Outline}

Offering unique operational fidelities and individual addressability, atomic
ions in radio-frequency (rf) traps are one of the most successful and
promising systems for quantum computation
\cite{wineland:qip:review, schindler:error:correction} and quantum metrology
\cite{chou:clockprl}. Due to strong short- as well as long-range interactions
in Coulomb crystals, they are also predestined for quantum simulation
experiments \cite{feynman:simulation} of, \eg, solid-state physics models
\cite{porras:quantum:spin:systems, friedenauer:quantum:magnet,
schneider:repprogphys}.
However, experiments on the quantum level with ions in rf traps have
been limited to the order of ten ions arranged in a linear string and
a common trapping potential \cite{islam:9ionsim, monz:14qubits}.
Experimental approaches to scaling particle numbers and
dimensionality of trapped-ion quantum simulations are mainly based on
surface-electrode micro-trap arrays
\cite{seidelin:trap, kumph:2d:arrays, clark:2dsimulations,
schneider:repprogphys} and Penning traps \cite{britton:ising}.

Extending the recent demonstration of ion trapping in a
single-beam dipole trap \cite{schneider:dipole:trap, schneider:dipole:trap:2} to
optical lattices has been proposed \cite{cirac:pushing:gate} to offer an
alternative route to scaling by combining the advantages of Coulomb interactions
with the scalability and versatility that have been developed for optical
lattices \cite{greiner:optical:lattices}. Such a system additionally allows for
storing ions and atoms in a common trap. This may become
essential \cite{cetina:fundamental:limit:ion:cooling} for ultracold atom--ion
collision experiments \cite{grier:collisions, zipkes:nature,
schmid:ion:in:bec, rellergert:atom:ion} because of the strong
suppression of micromotion \cite{cormick:optical:ion:trapping}. In this context
optical lattices may be useful, be it to increase trap depths, store several
ions/atoms in separate micro-wells or as conveyor belts
\cite{kuhr:conveyor:belt} for individual ions/atoms.

In the past, standing waves were already used in combination with ions and
rf traps to study particle dynamics \cite{katori:walther:lattice} and were
considered for preparing non-classical motional states
\cite{cirac:nonclassical:states} as well as implementing forces that depend on
the electronic state \cite{porras:quantum:spin:systems, cirac:pushing:gate}.
Additionally, there are proposals for quantum simulations requiring the local
shaping of the trapping potential of a rf-trapped Coulomb crystal by an optical
lattice \cite{schmied:optical:lattice, pruttivarasin:optical:lattice}.

\figurehandler{\myfigureschematic}

Here we report on trapping single ions in an all-optical trap,
where the confinement along the laser beam direction is provided by an optical
lattice, while the rf trap is switched off.

\figref{setup} shows a schematic of our setup. 
The experiments start by initializing single
$\atom[24]{Mg}[+]$ ions (nuclear spin $I=0$) in a linear rf
trap ($\omega_{\text{rf}}=\unit[2\pi\times 56]{MHz}$) \cite{schaetz:towards}.
This includes the creation of an ion by photo-ionization from a thermal atomic
beam and Doppler cooling to a few $\unit{mK}$ (Doppler cooling limit:
$\unit[1]{mK}$).
Initially, the oscillation frequencies of the ions in the ponderomotive, \ie{}
time-averaged, potential of the rf trap are
$\omega_{x,y}\approx\unit[2\pi\times
860]{kHz}$ radially and $\omega_{z}\approx\unit[2\pi\times 110]{kHz}$ axially.
 
Two dipole trap beams are arranged in a counter-propagating configuration,
providing light at a wavelength of $\lambda =\unit[280]{nm}$ and a power
$P_{\text{dip}}\leq\unit[100]{mW}$ in each beam \cite{friedenauer:laser}.
The dipole trap beams are red detuned ($\Delta_{\text{dip}}\approx
-\unit[2\pi\times 290]{GHz}$) from the $\level{S}[1/2]$--$\level{P}[3/2]$
transition.
The polarization of the beams can either be tuned to $\{ \sigma_{1}^+\text{,
}\sigma_{2}^-\}$, denoting the non-interfering configuration where beam $\# 1$
has $\sigma^+$ and beam $\# 2$ $\sigma^-$ polarization, or both beams are
$\sigma^+$ polarized, $\{\sigma_{1}^+\text{, }\sigma_{2}^+\}$, which allows for
their interference.
Coupling the $\level{S}[1/2]$ states to the $\level{P}[3/2]$
multiplett by the different polarization configurations relates to different
light shifts and, thus, dipole potentials of different depths for identical
laser intensities.
To permit a direct comparison between the two configurations we consider the
saturation parameter, 
$$s_{\text{dip}}=c\cdot
\frac{I_{\text{dip}}/I_{\text{sat}}}{1+(2\Delta_{\text{dip}} /\Gamma)^2}
\leq 3\times
10^{-3}\text{,}$$
with a linewidth $\Gamma =\unit[2\pi\times
41.8]{MHz}$, a saturation
intensity $I_{\text{sat}}=\unit[250]{mW/cm^2}$, and the single-beam
intensity $I_{\text{dip}}$.
The coupling strength is $c=1$ for
$\{\sigma_{1}^+\text{, }\sigma_{2}^+\}$ and $c=2/3$ for $\{
\sigma_{1}^+\text{,}\sigma_{2}^-\}$.

In the first stage of the experiment we compare the light shifts induced by a
single Gaussian laser beam, $\{ \sigma_{1}^+\}$, with that of two
counter-propagating beams, $\{\sigma_{1}^+\text{, }\sigma_{2}^+\}$, in order to
calibrate the interference of the dipole trap beams and to obtain estimates on
the relevant experimental imperfections.
We trap an ion in the rf trap and simultaneously induce a light shift.
We then detect the fluorescence of the ion induced by an additional low-power
probe beam as a function of $s_{\text{dip}}$ of the dipole trap beam(s).
The probe beam is $\sigma^+$ polarized and has an on-resonance saturation
parameter $s_0\approx 0.3$. It is overlapped with one of the dipole trap beams
and is (blue) detuned by $\Delta_{\text{probe}} = \unit[2 \pi \times
455]{MHz}\approx 10\Gamma$ with
respect to the unshifted $\level{S}[1/2]$--$\level{P}[3/2]$ transition. Thus,
resonance with the probe laser occurs when the light shift induced by the
dipole trap beam(s) matches the detuning of the probe laser. With
all lasers $\sigma^+$ polarized, we drive the closed cycling transition
$\ket{\level[][2]{S}[1/2],m_J=1/2}\leftrightarrow\ket{\level[][2]{P}[3/2],
m_J=3/2}$ and the relevant energy levels reduce to a two-level system.

\figurehandler{\myfiguress}

The results are shown in \figref{stark}.
For zero $s_{\text{dip}}$, the probe laser is far detuned, causing a
negligible fluorescence rate. Increasing the power of the dipole trap beam(s)
and, thus, $s_{\text{dip}}$, provides an increasing light shift. The electronic
transition approaches resonance with the probe laser, resulting in an increasing
fluorescence rate. The measurement with a single dipole trap beam serves as a
reference.
Its maximal fluorescence rate occurs at $s_{\text{dip}}\approx 1.28\times
10^{-3}$. In the measurement with $\{\sigma_{1}^+\text{,
}\sigma_{2}^+\}$ we find the resonance at lower $s_{\text{dip}}$ since less
single-beam power is needed to shift the transition into resonance with the
probe laser. The resonance is reduced in amplitude by
a factor of three and broadened towards higher $s_{\text{dip}}$.

We compare the experimental results with Monte Carlo
simulations (MCS) treating laser--ion interaction in rate-equations
\cite{schneider:dipole:trap:2} and relying on the ponderomotive
approximation for the rf trap potential. With the assumption of a thermal
initialization of the ions at $T_0=\unit[4]{mK}$
\footnotemark[1]\footnotetext{Further MCS parameters are the optimized
parameters of the rf trap potential from \refcite{schneider:dipole:trap:2},
$\alpha =\unit[35]{\degree}$, $\omega_{y'}^2 =- (2 \pi \times
\unit[50]{kHz})^2$, and $\omega_{x'} - \omega_{y'}=2 \pi
\times\unit[2.50]{kHz}$.}, the simulation is in good agreement with the
measurement results.
This confirms the formation of a standing wave at the position of the ion. 
The shapes and amplitudes of the resonances can be
explained by the oscillations of the ion in the combined trap consisting of the
rf and the optical potential. In particular, the broadened resonance in the
standing-wave case as well as the shift of the fluorescence maximum to
$s_{\text{dip}}\approx 0.37\times 10^{-3}$ in the MCS,
compared to its ideal location at a quarter of the single-beam resonance,
$s_{\text{dip}}\approx 1/4\cdot 1.28\times 10^{-3}=0.32\times 10^{-3}$, 
are due to the spatially averaged, and therefore reduced, light shift which the
oscillating ion experiences.
The experimental resonance in the standing-wave case occurs at
$s_{\text{dip}}\approx 0.42\times 10^{-3}$. This additional shift of the
saturation parameter scale by $\approx\unit[12]{\%}$ (see dashed line in
\figref{stark}) hints at an imperfect overlap of the dipole trap beams due
to, \eg, beam-pointing instabilities, corresponding to an average displacement
of the two beams by $\approx\unit[1]{\micro m}$. We use this result as an input
for the MCSs of the optical trapping experiments discussed in the following.

In the next stage of the experiment, we switch off the rf confinement completely
and measure the trapping probability for a constant trapping time as a function
of $s_{\text{dip}}$ in optical traps formed by counter-propagating dipole trap
beams, either interfering, $\{\sigma_{1}^+\text{, }\sigma_{2}^+\}$, or
non-interfering, $\{\sigma_{1}^+\text{, }\sigma_{2}^-\}$.

The experimental sequence is the following:
In order to optimize the transfer between rf trap and optical trap, we first
carefully compensate stray electric fields (for a quantitative estimate see
below). This is done by ramping down the rf potential to
$\omega_{x\text{,}y}\approx 2 \pi\times\unit[100]{kHz}$ and counteracting
the displacement of the ion with appropriate dc voltages
\cite{schneider:dipole:trap:2}.
For the optical trapping attempts, we ramp up the two
counter-propagating dipole trap beams and, subsequently, ramp the rf potential
down to zero, each in $\unit[50]{\micro s}$. After the optical trapping time
$T_{\text{opt}}$, the transfer protocol is reversed and the ion is detected
via resonance fluorescence in case of successful optical trapping.
During all steps a static electric potential in $z$ direction ($\omega_{z}=2
\pi\times\unit[45]{kHz}$) is retained, such
that, in the non-interfering case, the total confinement is due to the
dipole plus the static electric potential.
Compared to the standing-wave confinement, the contribution of the static
electric potential remains negligible.
According to Laplace's equation, its focussing effect along the $z$
axis even comes at the price of a
defocussing effect in at least one radial direction, which has to
be overcome by the optical potentials \cite{schneider:dipole:trap:2}.

\figurehandler{\myfiguretrapping}

\figref{trapping1} shows the trapping results for
$T_{\text{opt}}=\unit[25]{\micro s}$. For both polarization
configurations, at $s_{\text{dip}}=0$, and thus zero optical
trap depth, the trapping probability is found to be zero, verifying that
after turning off the rf trap there remain no significant residual trapping
potentials.
In the non-interfering case the optical trapping probability rises with
increasing $s_{\text{dip}}$ and reaches close to $P=\unit[100]{\%}$.
In the standing-wave case, the trapping probability exceeds that of the
non-interfering case for $s_{\text{dip}}\lesssim 0.6\times 10^{-3}$, but
then levels off at $P\approx\unit[80]{\%}$.

The gradual rise of optical trapping probability with increasing
$s_{\text{dip}}$ is due to the non-zero initial temperature.
For the standing-wave case, the trapping probability rises faster due to the
trap depth being increased, ideally, by a factor of two. 
The reduced trapping probability for $s_{\text{dip}} > 1\times 10^{-3}$
cannot be explained by laser-induced heating effects because these should remain
negligible whithin this regime \cite{gordon:ashkin} (the explanation
follows in the context of \figref{trapping2}).

Shown in \figref{trapping1} are also the results of the MCSs. These incorporate
the full transfer sequence of the ion from the ponderomotive potential of the rf
trap into the dipole trap (as in \refcite{schneider:dipole:trap:2}) as well as
dressed-state rate equations \cite{dalibard:dressed:states}, which allow to
reproduce both recoil heating and dipole-force fluctuation heating
\cite{gordon:ashkin}.

In the non-interfering case, good agreement with the measurements is
obtained, again assuming $T_0 = \unit[4]{mK}$. The small trapping probability at
$s_{\text{dip}}\approx 0.4\times 10^{-3}$ is reproduced assuming a
constant force $F=\unit[0.5\times 10^{-20}]{N}$ radial to the
trapping beam. This gives an estimate for the limitation of our
stray-field compensation procedure and matches very well
the rough calculation of $F=\unit[10^{-20}]{N}$ which we made in
\refcite{schneider:dipole:trap}.
The simulation in the standing-wave case, again with $T_0 =\unit[4]{mK}$,
reproduces the faster rise at small $s_{\text{dip}}$. However, it also
predicts $P\approx\unit[100]{\%}$ for $s_{\text{dip}}>1.5\times 10^{-3}$, which
is not observed in the experiment.

Remarkably the harmonically approximated oscillation frequency along the
standing-wave direction reaches $\omega_{\text{dip}}\approx\unit[2\pi\times
30]{MHz}\approx\omega_{\text{rf}}/2$ at $s_{\text{dip}}\approx 3\times
10^{-3}$.
To reveal the causal link between the reduced trapping probabilities
and the rf field, we modify the previous (direct transfer) protocol of loading
the ion from the rf trap into the standing-wave trap:
We insert the transfer into a single-beam optical trap and
ramp up the second dipole trap beam only after the rf amplitude has been ramped
down to zero (indirect transfer). Thus, we avoid the temporal overlap of
standing wave and rf field. 

\figurehandler{\myfiguretrappingrf}

The results for single-beam saturation parameters of
$s_{\text{dip}}\approx 1.5\times 10^{-3}$ and $s_{\text{dip}}\approx
2.8\times 10^{-3}$ are shown in \figref{trapping2}. Also depicted are the
experimental results for the direct transfer process from \figref{trapping1}
with the extended range of $s_{\text{dip}}\leq 3\times 10^{-3}$, where for
$s_{\text{dip}}>1.5\times 10^{-3}$ the increase of $s_{\text{dip}}$ leads to a
decrease
of the trapping probability to below $\unit[60]{\%}$. In contrast, with the
indirect transfer, we significantly increase the trapping probability, \eg{}
by $\unit[30]{\%}$ at $s_{\text{dip}}=2.8\times 10^{-3}$. Still, these results
represent a lower bound to the actual improvement since additional losses
occur during the single-beam trapping phase with $s_{\text{dip}}\approx
2.8\times 10^{-3}$ and the increase of the optical trapping time to
$T_{\text{opt}}=\unit[100]{\micro s}$, as required for the extended protocol.

\figref{trapping2} also shows results of MCSs that incorporate the
time-dependent rf potential. For the direct transfer, the drop in trapping
probability for $1.5\times 10^{-3}<s_{\text{dip}}<3\times 10^{-3}$ is
reproduced well. For the indirect transfer we retrieve the significantly
increased trapping probabilities.
This is evidence that the observed reduction in trapping probability
for the direct transfer is mainly due to excitation of the ion motion within the
stiff potential of the standing wave by the rf driving field:
$\omega_{\text{dip}}$ approaching $\omega_{\text{rf}}/2$ leads to
increasing parametric excitations.
This may also explain results we obtained with hybrid traps
consisting of the rf potential and red- or blue-detuned standing waves, where
the optical trapping probability deteriorated compared to the exclusively
optical confinement described above. 
For both transfer protocols of \figref{trapping2}, the observed trapping
probabilities remain below the simulation results. This, presumably, is a result
of additional experimental imperfections. For the direct loading, for example,
spatially not perfectly overlapped traps (offset $\vec{d}$) lead to additonal
micromotion during transfer, which amplifies the parametric excitation. In fact,
by adapting the MCS parameters to $F=\unit[1.5\times 10^{-20}]{N}$ and
$\vec{d}=\unit[0.35]{\micro m}\times\vec{e}_x$, very good agreement
between simulation and experiment can be reached (gray lines in
\figref{trapping2}).
 
Our results might have to be considered in future experiments, in particular for
hybrid (rf/optical) traps such as those currently used \cite{zipkes:nature,
schmid:ion:in:bec, rellergert:atom:ion} and proposed
\cite{cetina:fundamental:limit:ion:cooling} for
ultra-cold atom--ion collision experiments or quantum simulations
\cite{pruttivarasin:optical:lattice, schmied:optical:lattice}. 
Apart from the method of indirect loading, demonstrated here,
possible ways to minimize rf heating of the ion are high radio frequencies, far
beyond the oscillation frequencies in the lattice, and/or low
oscillation frequencies in the lattice due to longer wavelengths. 
A standing wave aligned with the rf trap axis could reduce rf-induced
parametric heating by minimizing the projection of rf forces on the
standing-wave direction. If a blue-detuned laser was used, recoil heating
could be suppressed as well. Enhanced laser intensities without
optical lattices and the pertinent heating effects could be achived with optical
ring resonators.
Apart from their application in hybrid traps, optical lattices for ions, or ions
and atoms, are themselves a promising system, for example to combine Coulomb or
charge-exchange interactions with scalability for quantum simulation
experiments as discussed in
\refsscite{cirac:pushing:gate}{schneider:dipole:trap}{schneider:repprogphys}.
To avoid temporal overlap between rf and optical lattice during loading from a
rf trap, turning off the rf drive faster would be advantageous. However, the
ring-down time of the rf resonance circuit has to be considered as well as
additional heating effects in case of non-adiabatic changes of the potentials.
The limitations of our experiment in terms of lifetimes in the optical ion traps
can be overcome by using high-power, far-off-resonance lasers, although the
need for high powers may be reduced by employing optical cavities.

% \iftwocolumn{}{\clearpage}

%\section*{Acknowledgement}

This work was supported by MPQ, MPG, DFG (SCHA
973/1-6), and EU (PICC, grant no. 249958). We thank Stephan D\"{u}rr for helpful
discussions and Govinda Clos and Julian Schmidt for comments on the
manuscript.

% This work was supported by the Max-Planck-Institut f\"{u}r Quantenoptik (MPQ),
% Max-Planck-Gesellschaft (MPG), Deutsche Forschungsgemeinschaft (DFG) (SCHA
% 973/1-6), the European Commission (The Physics of Ion Coulomb Crystals: FP7
% 2007--2013. grant no. 249958) and the DFG Cluster of Excellence ``Munich
% Centre for Advanced Photonics''. 
% 

\iftwocolumn{}{\clearpage}

\bibliographystyle{\mybibstyle}

%Merlin.mbs v4.21 2009-07-09.
%

%\bibliography{references}
%\bibliographystyle{\mybibstyle}

\iftwocolumn{}{\clearpage}

\thefigurepages

\therevisionchanges

\end{document}